\documentclass[fleqn,usenatbib]{mnras}

\usepackage{newtxtext,newtxmath}

\usepackage[T1]{fontenc}

\DeclareRobustCommand{\VAN}[3]{#2}
\let\VANthebibliography\thebibliography
\def\thebibliography{\DeclareRobustCommand{\VAN}[3]{##3}\VANthebibliography}

\usepackage{graphicx}	
\usepackage{amsmath}	

\title[A radio bridge in Abell 85]{A radio bridge connecting the minihalo and phoenix in the Abell 85 cluster}

\author[Ramij Raja et al.]{
Ramij Raja,$^{1}$\thanks{E-mail: ramij.amu48@gmail.com}
Majidul Rahaman,$^{2}$
Abhirup Datta$^{3}$
and Oleg M. Smirnov$^{1,4}$
\\
$^{1}$Centre for Radio Astronomy Techniques and Technologies, Department of Physics and Electronics, Rhodes University, Makhanda 6140, South Africa\\
$^{2}$Institute of Astronomy, National Tsing Hua University, Hsinchu 300044, Taiwan, R.O.C.\\
$^{3}$Department of Astronomy, Astrophysics and Space Engineering, Indian Institute of Technology Indore, Simrol, 453552, India \\
$^{4}$South African Radio Astronomy Observatory, 2 Fir Street, Black River Park, Observatory, Cape Town 7925, South Africa
}

\date{Accepted XXX. Received YYY; in original form ZZZ}

\pubyear{2023}

\begin{document}
\label{firstpage}
\pagerange{\pageref{firstpage}--\pageref{lastpage}}
\maketitle

\begin{abstract}
Galaxy clusters are located at the nodes of cosmic filaments and therefore host a lot of hydrodynamical activity. However, cool core clusters are considered to be relatively relaxed systems without much merging activity. The Abell 85 cluster is a unique example where the cluster hosts both a cool core and multiple ongoing merging processes. 
In this work, we used 700 MHz uGMRT as well as MeerKAT L-band observations, carried out as part of the MGCLS, of the Abell 85. 
We reconfirm the presence of a minihalo in the cluster centre at 700MHz that was recently discovered in MGCLS. Furthermore, we discovered a radio bridge connecting the central minihalo and the peripheral radio phoenix.
The mean surface brightness, size and flux density of the bridge at 700 MHz is found to be $\sim 0.14\ \mu$Jy/arcsec$^2$, $\sim 220$ kpc and $\sim 4.88$ mJy, respectively, with a spectral index of $\alpha_{700}^{1.28} = -0.92$. Although the origin of the seed relativistic electrons is still unknown, turbulent re-acceleration caused by both the spiralling sloshing gas in the intracluster medium (ICM) and the post-shock turbulence from the outgoing merging shock associated with the phoenix formation may be responsible for the bridge.

\end{abstract}




\section{Introduction}
The revolution in sensitive low-frequency observations with modern telescopes such as LOFAR, uGMRT, MeerKAT etc. has broadened our horizons on the presence of diffuse non-thermal emission associated with clusters of galaxies. Apart from the classical radio halos, minihalos, relics and lately the radio phoenices \citep{Kempner2004rcfg.proc..335K} from the fossil plasma, we have started discovering another new class of diffuse radio object known as the radio bridge \citep[see][for latest review]{vanWeeren2019SSRv..215...16V}.
After the first reported radio bridge in the Coma-A1367 supercluster \citep{Kim1989Natur.341..720K}, a variety of radio bridges were discovered in different scenarios, from a bridge connecting halo and relic \citep[e.g., 1RXSJ0603.3+4214,][]{vanWeeren2012A&A...546A.124V,Rajpurohit2018ApJ...852...65R} to a giant bridge ($\sim 3$ Mpc) between two pre-merging cluster systems \citep[e.g., A399-401, A1758,][]{Govoni2019Sci...364..981G,Botteon2020MNRAS.499L..11B}. In this work, we report the discovery of another radio bridge connecting the minihalo at the centre and the radio phoenix at the periphery of the Abell 85 cluster. 

The Abell 85 (hereafter A85) cluster is a complex system with a cool core and ongoing merging activity. This cluster is situated at a redshift $z=0.
0556$ \citep{Rines2016ApJ...819...63R} and hosts two ongoing subcluster mergers from south-west and southern direction \citep{Kempner2002ApJ...579..236K,Durret2005A&A...432..809D} as well as multiple galaxy group mergers \citet{Yu2016ApJ...831..156Y}, which makes this cluster one of the best candidates to study the ICM dynamical processes. Being such a case, quite a few multi-frequency radio investigations were reported in the literature on this cluster; targeting mainly the radio phoenix \citep{Kempner2004rcfg.proc..335K} situated at the south-western periphery of the cluster \citealt{Slee1984PASAu...5..516S,Bagchi1998MNRAS.296L..23B,Giovannini1999NewA....4..141G,Giovannini2000NewA....5..335G,Slee2001AJ....122.1172S,Duchesne2021PASA...38...10D,Rahaman2022MNRAS.515.2245R}. Although \citet{Ichinohe2015MNRAS.448.2971I} reported the presence of gas sloshing in the ICM, no minihalo emission was detected in any of the previous studies. 
Only recently, with the help of the powerful MeerKAT telescope, an elongated diffuse emission at the centre around the Brightest Cluster Galaxy (BCG) was finally discovered at 1.28 GHz by \citet{Knowles2022A&A...657A..56K} in the MeerKAT Galaxy Cluster Legacy Survey (MGCLS) and was classified as a minihalo. However, they have also commented that the elongated nature of the minihalo, along with a central radio-loud BCG was found in other clusters as well in their survey, and they might possibly be a new kind of radio halo. However, with sensitive low-frequency uGMRT observations, we have discovered a radio bridge along the elongation which is discussed in detail below.

In this Letter, we report the discovery of a radio bridge connecting the radio minihalo and the radio phoenix in the A85 cluster. Here, we adopt a $\Lambda$CDM cosmology with $H_0 = 70$ km s$^{-1}$ Mpc$^{-1}$, $\Omega_{\mathrm{m}} = 0.3$ and $\Omega_\Lambda = 0.7$. At the cluster redshift $z = 0.0556$, $1\arcsec$ corresponds to a physical scale of $1.08$ kpc.

\section{Observations and Data Analysis} \label{sect:obs}
\subsection{uGMRT Band-4 data}
In this study, we have used our recent observation of the A85 cluster with the uGMRT in band-4. The observation was made in the frequency range of $550 - 850$ MHz with a usable bandwidth of $\sim 200$ MHz, divided into 8k channels. The total synthesis time is $\sim 11$ hrs on 20 Nov 2020 (Project code: 39\_104, PI: Ramij Raja). 

For data reduction, we used the updated SPAM package \citep{Intema2009A&A...501.1185I,Intema2017A&A...598A..78I} which can process the wideband uGMRT data. First, we split the raw FITS data into chunks by splitting the wide bandwidth into several subbands using the task \textit{split\_wideband\_uvdata}. Next, each subband data is processed independently starting with the task \textit{pre\_calibrate\_wideband\_targets} where calibration and flagging information are derived for the primary calibrator and then transferred to the target data. However, one needs to set a single reference frequency for all the subbands at this stage in order to combine all the calibrated visibility chunks later and produce an MFS (Multi Frequency Synthesis) image. After that, the pre-calibrated target visibilities were processed in the main pipeline \textit{process\_wideband\_target} where repeated flagging, self-calibration and finally direction-dependent calibration were performed. In this stage, as an initial sky model, we provided a source catalogue derived using PyBDSF \citep{Mohan2015ascl.soft02007M} from the SPAM output image of the narrow-band GMRT (GSB) data. For a more detailed description of each step, please refer to the SPAM webpage\footnote{http://www.intema.nl/doku.php?id=huibintemaspampipeline}. Afterwards, the calibrated visibilities for each subband in UVFITS format are collected in a single directory and converted to MS (Measurement Set) format with the CASA \citep{McMullin2007ASPC..376..127M} task \textit{importuvfits} for use in later steps. Finally, the desired images were produced by combining all the calibrated subband MS using WSCLEAN \citep{offringa-wsclean-2014,Offringa2017MNRAS.471..301O}. The imaging parameters used are described as corresponding to the specific images discussed below. 

\subsection{MeerKAT L-band data}
The MeerKAT L-band data of the A85 cluster used in this work was observed as a part of the MGCLS survey \citep{Knowles2022A&A...657A..56K}. The A85 cluster was observed on 26 Sept 2018 using the full MeerKAT array at the frequency band $900-1670$ MHz in the 4k correlator mode. The integration time was 8 sec with a total observing time of $\sim 8$ hrs (Project ID: SSV-20180624-FC-01).
For data reduction, we used the CARACal pipeline \citep{Jozsa2020} which employs multiple packages like CASA, TRICOLOUR, WSCLEAN etc. for data editing, calibration and imaging. 
Using this pipeline we performed basic radio interferometric data reduction steps to get the calibrated target data. After that, we performed a few rounds of phase-only self-calibration to remove residual phase errors. For a more detailed description of each step see the CARACal webpage\footnote{https://caracal.readthedocs.io/en/latest/}. 
\section{Results} \label{sect:res}

\begin{figure}
    \centering
    \includegraphics[width=0.99\columnwidth]{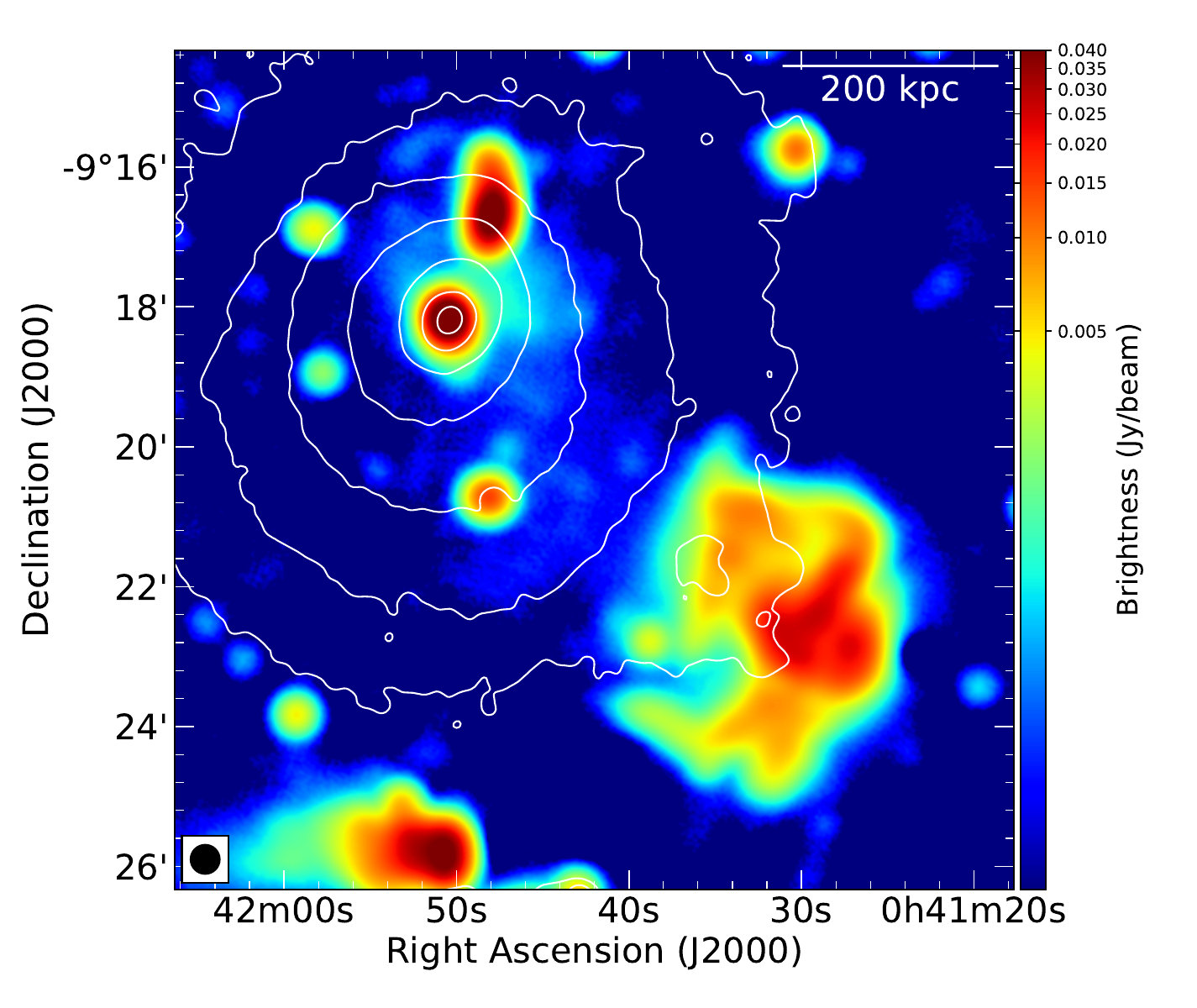}
    \caption{Low-resolution GMRT radio map of the A85 at 700 MHz. The map noise and restoring beam of this image are $\sigma_{\mathrm{rms}}^{700}$ = 23 $\mu$Jy beam$^{-1}$ and $\mathrm{Beam_{700}} = 25\arcsec$, respectively. The X-ray brightness distribution in the cluster is shown with white contours.} 
    \label{fig:radio_xray}
\end{figure}

\begin{figure*}
    \centering
    \begin{tabular}{cc}
    \includegraphics[width=0.99\columnwidth]{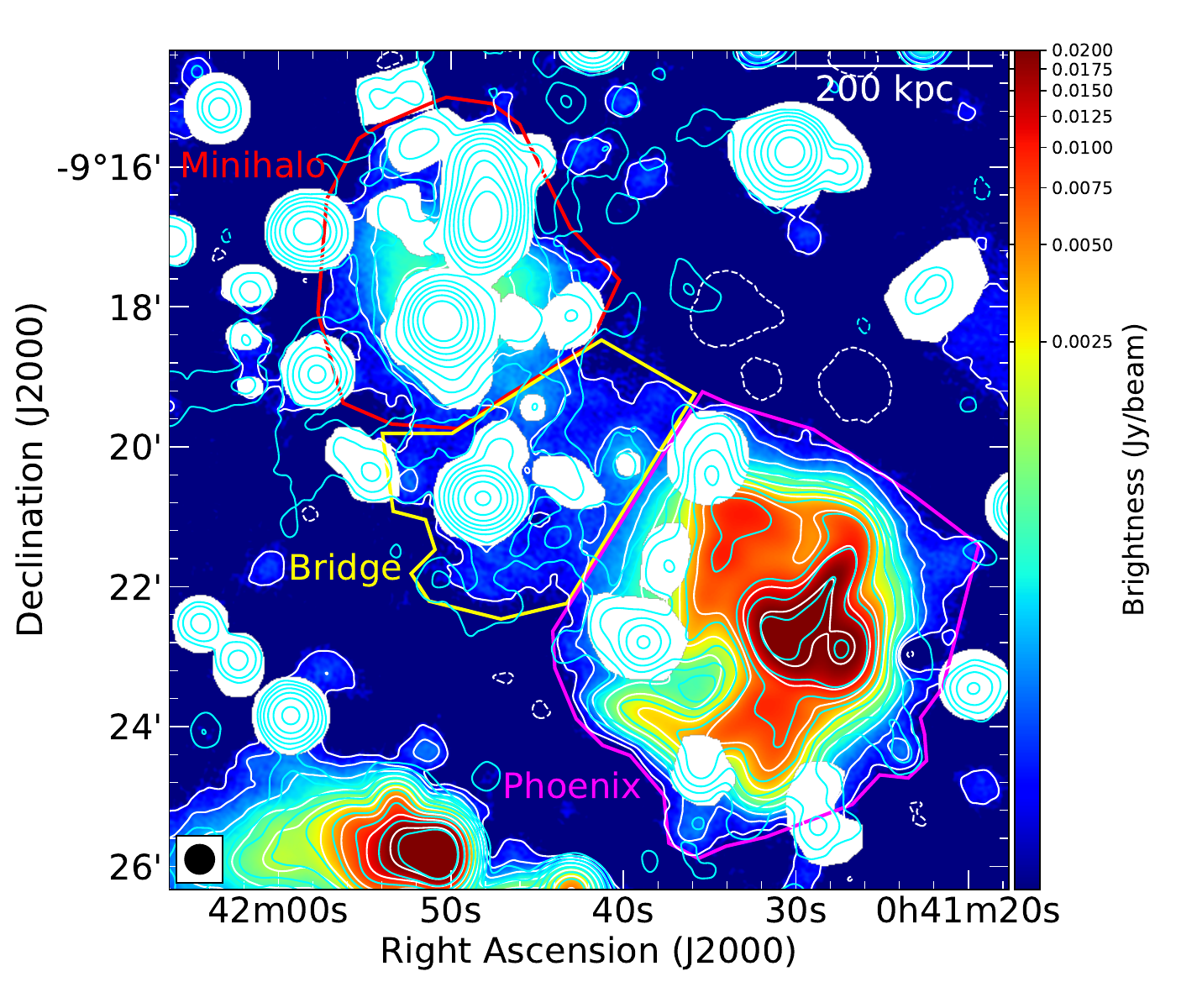} &
    \includegraphics[width=0.99\columnwidth]{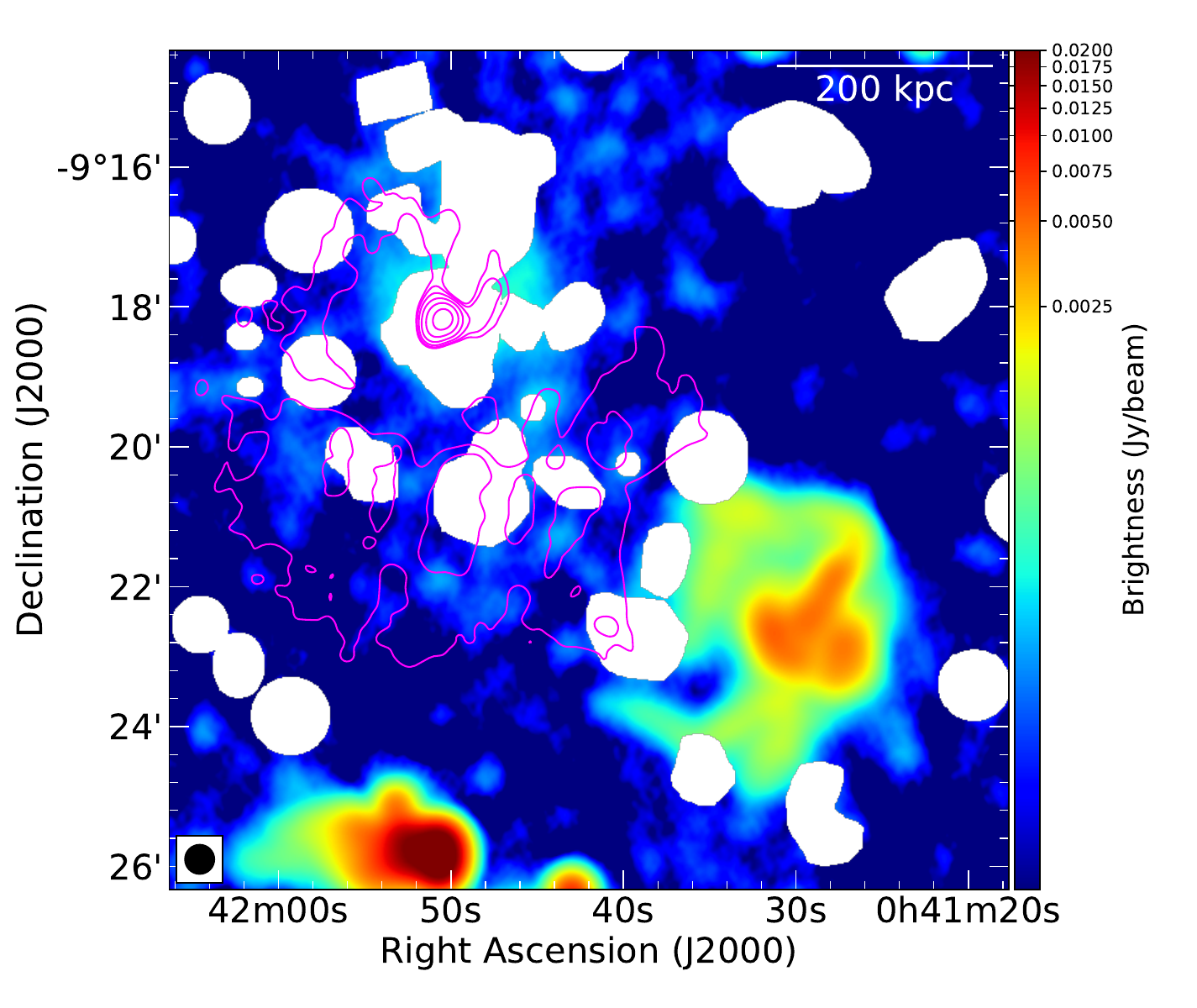} \\
    \end{tabular}
    \caption{\textit{Left}: Radio galaxy masked uGMRT 700 MHz image of the A85 cluster overlaid with both 700 and 1280 MHz image (unmasked) contours. The radio contours are drawn at levels $[-1,1,2,4,8,...]\times 3\sigma_{\mathrm{rms}}$. The map noise of these images are $\sigma_{\mathrm{rms}}^{700}$ = 27 $\mu$Jy beam$^{-1}$ and $\sigma_{\mathrm{rms}}^{1280}$ = 22 $\mu$Jy beam$^{-1}$, respectively. The restoring beam of these images is $25\arcsec$.  \textit{Right}: MeerKAT 1280 MHz radio galaxy masked image overlaid with X-ray residual brightness contours.} 
    \label{fig:uvsub_main}
\end{figure*}

In Fig. \ref{fig:radio_xray}, the \textit{Chandra} X-ray brightness contours show both the extent of the ICM as well as the relaxed morphology (observed in cool core clusters) with near concentric brightness contours (see \citealt{Rahaman2022MNRAS.515.2245R} for more details). In the background, the deep uGMRT image of the A85 cluster at 700 MHz is presented. This image was produced using full visibilities with Briggs robust = 0 \citep{Briggs1995}. To amplify the diffuse radio emission, we restored the image with a $25\arcsec$ beam. In this image, the well-known radio phoenix \citep[e.g.,][]{Giovannini2000NewA....5..335G} at the cluster periphery, the recently discovered radio minihalo at the cluster centre \citep{Knowles2022A&A...657A..56K} and a new radio bridge connecting both of them are clearly detected. We also tried tapering the longer baselines, but the resulting images only lost sensitivity without revealing any more diffuse emission. 

The images used to derive results in this work at both 700 and 1280 MHz frequencies were produced by selecting a common \textit{uv}-range ($0.1-43$k$\lambda$) with Briggs robust = 0 and finally restored with $25\arcsec$ beam. These images are presented in Fig. \ref{fig:uvsub_main}. For quantitative analysis of the images, a proper subtraction of the embedded radio galaxies and a `zero level correction' of the images is necessary. Since perfect subtraction of the radio galaxy contributions is extremely difficult, we proceeded with masking the affected regions. Furthermore, since we are more interested in the spectral nature of the diffuse objects rather than their absolute integrated flux density values, this approach ensures the complete exclusion of radio galaxy flux density contributions in our analysis. Next, `local RMS' and `zero level' were determined and corrected for both 700 MHz and 1.28 GHz images. A detailed description of the process is presented in Appendix \ref{sec:zero_level}. It should be noted that all results presented in this work are derived from the `zero level' corrected, radio galaxy-masked images at both frequencies (Fig. \ref{fig:uvsub_main}).

For clarity, we have labelled the rough extent of the minihalo, bridge and phoenix in Fig. \ref{fig:uvsub_main} (left). Furthermore, the flux density values of the diffuse objects are estimated at both frequencies within the indicated regions. The integrated flux densities of the minihalo, bridge and phoenix at 700 MHz are estimated to be $6.92\pm0.69$, $4.88\pm0.69$ and $475.49\pm28.58$ mJy, respectively. At 1.28 GHz, the estimated flux densities of the minihalo, bridge and phoenix are found to be $3.85\pm0.55$, $2.80\pm0.59$ and $77.8\pm4.21$ mJy, respectively. It should be noted that these flux density values are under-estimated as the masked regions were excluded. 
The uncertainty in the flux density values is derived using

\begin{equation}
    \sigma_{S} = \sqrt{(\sigma_{\mathrm{cal}}S)^2 + (\sigma_{\mathrm{rms}}\sqrt{N_{\mathrm{beam}}})^2 + (zls*N_{\mathrm{beam}})^2}\,,
    \label{eq:flux_err}
\end{equation}

\noindent where $S$ is flux density, $\sigma_{\mathrm{cal}}$ and $\sigma_{\mathrm{rms}}$ are calibration uncertainty and map noise, respectively, $N_{\mathrm{beam}}$ is number of beams and $zls$ is `zero level scatter' (See Appendix. \ref{sec:zero_level}). We assume an absolute flux calibration error of $6\%$ \citep{Chandra2004ApJ...612..974C} for 700 MHz and $5\%$ for 1.28 GHz. We found the integrated spectral index of the minihalo, bridge and phoenix to be $-0.97\pm0.29$, $-0.92\pm0.42$ and $-3.00\pm0.13$, respectively. We have skipped the analysis of the phoenix here, as we have recently performed a detailed multi-frequency study of the radio phoenix in Raja et al. (in prep.). 

\begin{figure}
    \centering
    \includegraphics[width=0.99\columnwidth]{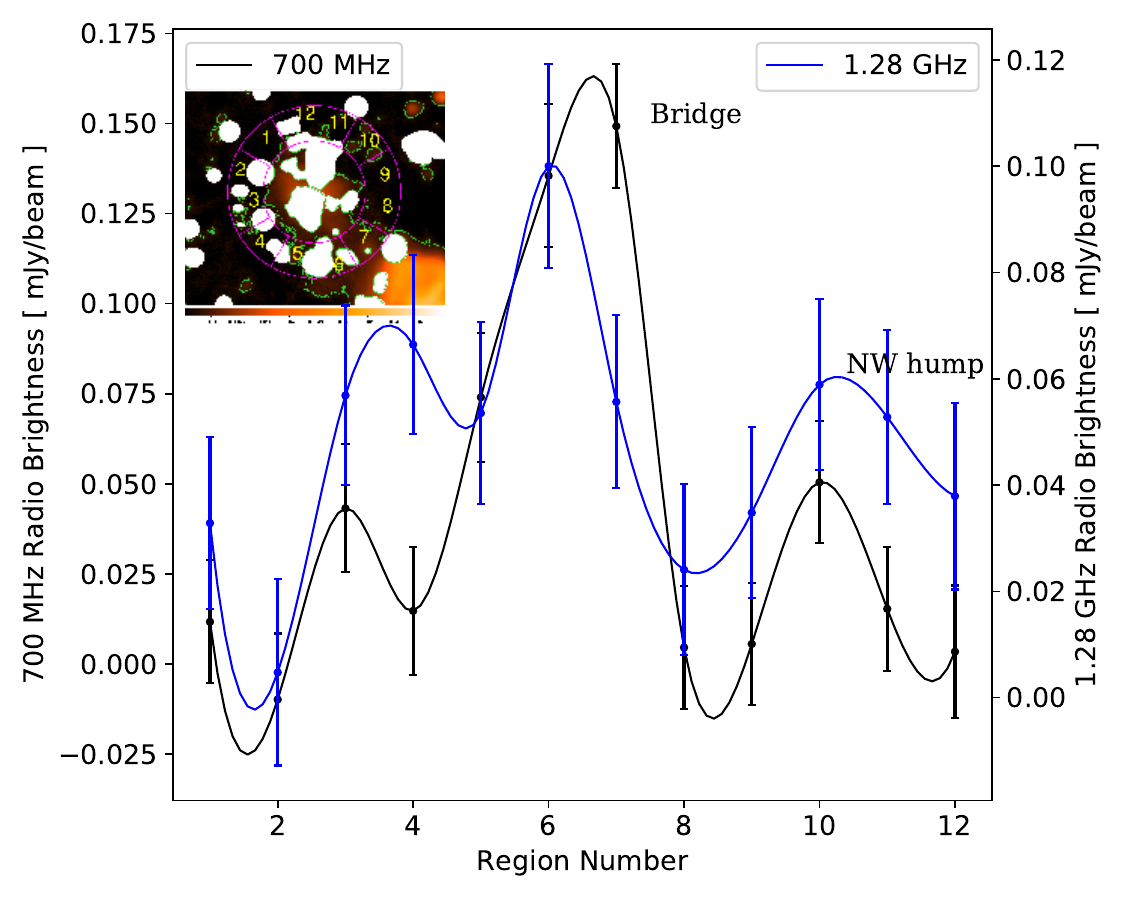}
    \caption{The azimuthal profile around the cluster as indicated in the inset image. The data points are the mean brightness value in each bin whereas the error bars indicate $\sqrt{(\sigma_\mathrm{rms}/\sqrt{N_\mathrm{beam}})^2 + zls^2}$, where $\sigma_\mathrm{rms}$ is the map noise, $N_\mathrm{beam}$ is the number of beams in each bin and $zls$ is `zero level scatter'. Radio galaxy locations are masked to avoid any contributions to the profile.}
    \label{fig:az_profile}
\end{figure}

\begin{figure*}
    \centering
    \begin{tabular}{cc}
    \includegraphics[width=0.99\columnwidth]{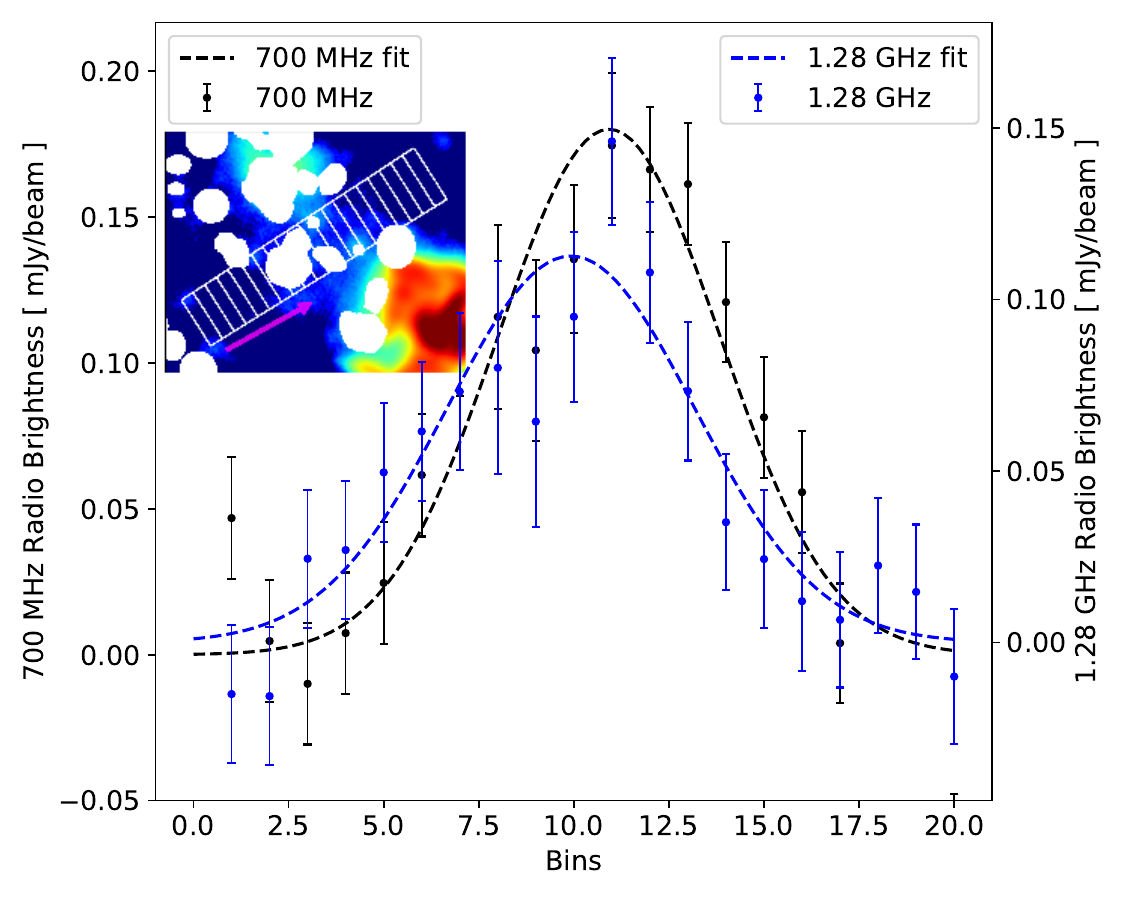} & 
    \includegraphics[width=0.99\columnwidth]{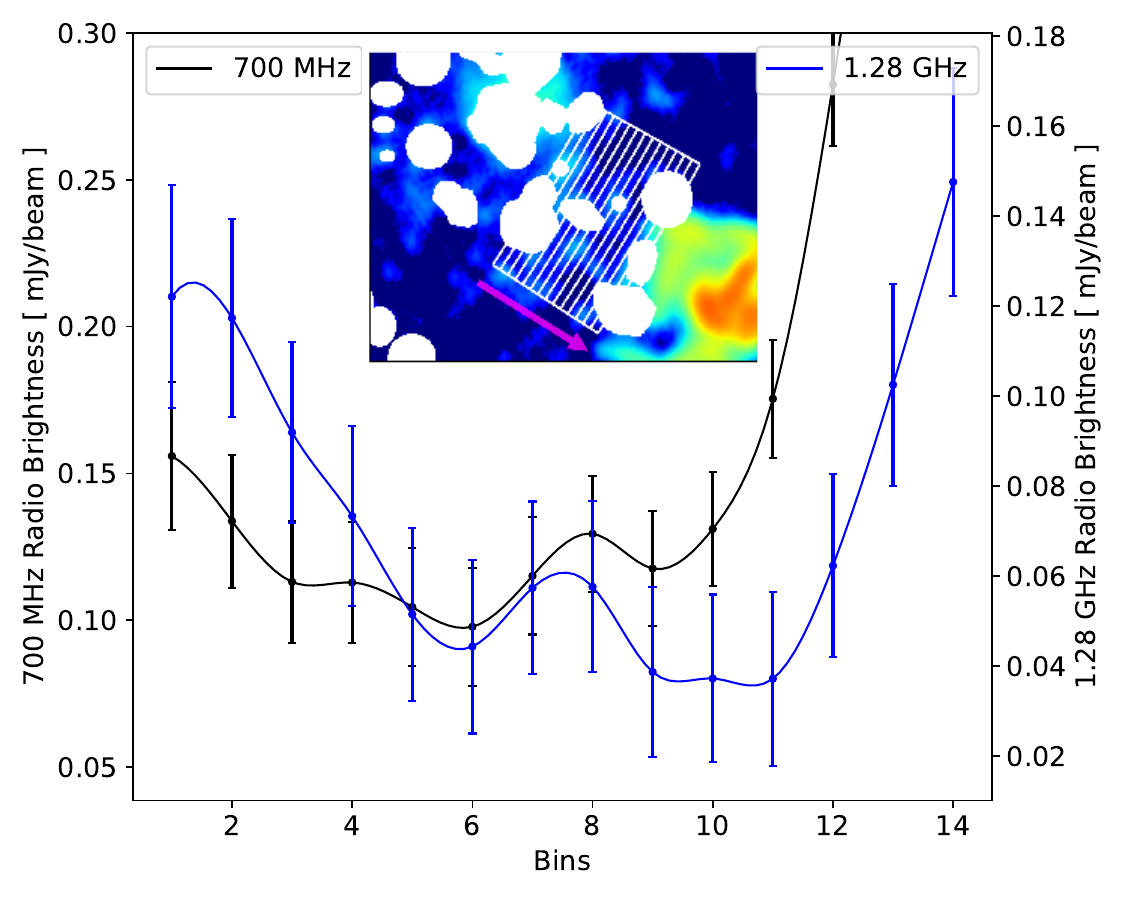} \\
    \end{tabular}
    \caption{\textit{Left}: The radio brightness profiles across the bridge as indicated in the inset image. The data points are the average brightness value in each bin whereas the error bars indicate $\sqrt{(\sigma_\mathrm{rms}/\sqrt{N_\mathrm{beam}})^2 + zls^2}$, where $\sigma_\mathrm{rms}$ is the map noise, $N_\mathrm{beam}$ is the number of beams in each bin and $zls$ is `zero level scatter'. The bin width is $25\arcsec$, the same as the restoring beam of the image. Dashed curves are the best fit Gaussian model corresponding to the data points. \textit{Right}: The radio brightness profile from the minihalo to the phoenix along the bridge as indicated in the inset image. This plot is produced the same way as the previous one with bin width $12.5\arcsec$. Radio galaxy locations are masked to avoid any contributions to the profiles. The magenta arrow indicates the profile direction.}
    \label{fig:bridge_profile}
\end{figure*}

\section{Discussion} \label{sect:disc}
\subsection{The minihalo} 

In this work, we reconfirm the presence of the minihalo at the centre of the A85 cluster recently discovered by \citet{Knowles2022A&A...657A..56K} with our 700 MHz uGMRT observation. The primary reason most of the previous investigations failed to detect this diffuse object is (a) the very low surface brightness ($\sim 0.4\ \mu$Jy/arcsec$^2$ at 700 MHz) and (b) the presence of the bright BCG and tailed radio galaxy. 
We estimate the size of the minihalo at 700 MHz to be $\sim 280$ kpc. 
The spectral index of the minihalo between 700 MHz and 1.28 GHz is found to be $-0.97\pm0.29$, which is around the typical minihalo spectral index ($\sim -1.2$; e.g., \citealt{Ferrari2011A&A...534L..12F,Giacintucci2011A&A...525L..10G,vanWeeren2019SSRv..215...16V}). 

As the presence of sloshing gas in the ICM is already reported by \citet{Ichinohe2015MNRAS.448.2971I}, the discovery of the minihalo at the centre of the A85 further supports the widely accepted minihalo formation mechanism proposed by \citet{Fujita2004ApJ...612L...9F} and later confirmed by  \citet{Mazzotta2008ApJ...675L...9M}. That is, the minor merger-induced sloshing ICM gas injects turbulence into the medium which is responsible for the re-acceleration of the \textit{in situ} seed relativistic electrons. 

In Fig. \ref{fig:uvsub_main} (left) we observed that the south-eastern part of the minihalo is not completely detected at 700 MHz. This is much more obvious in the azimuthal profile presented in Fig. \ref{fig:az_profile}. Here, we notice that the brightness in Region 4 is less compared to adjacent regions at 700 MHz, whereas at 1.28 GHz the brightness remains relatively similar. Furthermore, we observe that the surface brightness of the outer part of the minihalo is almost half as much as that of the bridge at 1.28 GHz. Therefore, it is difficult to detect the outer part of the minihalo at 700 MHz, as the bridge itself is recovered only at around $3-4\sigma$ level. Furthermore, the short baselines ($<1$k$\lambda$) at the 700 MHz GMRT observation are not as well sampled as in the 1.28 GHz MeerKAT observation (see Fig. \ref{fig:uvplot}). This further highlights the capability of the MeerKAT array in detecting faint large-scale diffuse radio emission.

\subsection{The radio bridge}

In Fig. \ref{fig:uvsub_main} (left), the extended diffuse emission present in the A85 cluster is visible at 700 MHz. 
Spanning from the minihalo at the centre with bright spiralling gas to the phoenix at the periphery with complex filamentary structure, the radio bridge is a stretch of diffuse radio emission of about $\sim 220$ kpc. The average brightness of the bridge at 700 MHz is $\sim 0.14\ \mu$Jy/arcsec$^2$, which is too faint to be detected by any previous observations. In the literature, it is observed that the size of radio bridges varies depending upon their formation scenario. A bridge between two pre-merging clusters spans around $2\sim3$ Mpc \citep{Govoni2019Sci...364..981G,Botteon2020MNRAS.499L..11B} wheres a bridge connecting a halo and relic is around $\sim600$ kpc \citep{Pasini2022A&A...663A.105P}. Furthermore, the brightness of the bridges found in the literature is mostly in the $\mu$Jy level as well e.g., $\sim 0.4$ \citep{Govoni2019Sci...364..981G}, $\sim0.2$ \citep{Botteon2020MNRAS.499L..11B}, $\sim0.6$ \citep{Bonafede2021ApJ...907...32B}, $\sim1$ \citep{Pasini2022A&A...663A.105P} at 144MHz and $\sim 0.25\ \mu$Jy/arcsec$^2$ \citep{Carretti2013MNRAS.430.1414C} at 1.4GHz. To emphasise that the bridge is not a slight enhancement over a more general symmetric minihalo, we have plotted an azimuthal profile around the outer region of the minihalo (Fig. \ref{fig:az_profile}). Here, the bridge is detected at both 700 MHz and 1.28 GHz at high significance compared to adjacent azimuthal bins.
Another interesting feature observed in the profile is the north-west hump at the wake of the outgoing radio source, which may also contribute to the formation of diffuse radio objects in the A85 cluster.
In Fig. \ref{fig:bridge_profile}, we have plotted the brightness profile of the bridge in two different directions. The left panel shows a profile across the bridge from southeast to northwest. We see the presence of the bridge with a significant increase in brightness profile compared to the map noise of the image.
A Gaussian model is fitted to the data to show the significance of the emission. The bridge emission peaks at $\sim 0.18$ mJy/beam with an FWHM of $\sim 185$ kpc at 700 MHz and $\sim 0.11$ mJy/beam with an FWHM of $\sim 207$ kpc at 1.28 GHz. The slightly wider fit at 1.28 GHz indicates that the bridge is detected at a lower significance compared to the 700 MHz observation. 
On the right panel, we plotted the brightness profile from the minihalo to the phoenix along the bridge. Here we see that the bridge brightness decreases along the bridge until the radio phoenix where again the brightness increases rapidly. This trend is observed at both 700 MHz and 1.28 GHz images.
The integrated spectral index of the bridge between these frequencies is found to be $-0.92\pm0.42$. This is much flatter compared to the spectral indices of $-1.3 \lesssim \alpha \lesssim -1.7$ for bridges found in the literature \citep{Carretti2013MNRAS.430.1414C,Botteon2020MNRAS.499L..11B,Bonafede2021ApJ...907...32B,Pasini2022A&A...663A.105P}, although that lower limit is compatible with our error bar.

So far, very few radio bridges in galaxy clusters have been discovered (e.g., 1RXSJ0603.3+4214, A399-401, A1758, A3667) and hence we have little knowledge about their origin. For Mpc scale inter-cluster bridges, \citet{Brunetti2020PhRvL.124e1101B} proposed a scenario where the complex dynamical activity in the overdense region between the clusters injects turbulence which may provide the necessary energy to observe this emission. On the other hand, the possible origin of the radio bridge connecting the halo and relic in the 1RXSJ0603.3+4214 was suggested by \citet{vanWeeren2012A&A...546A.124V} to be the post-shock turbulence behind the relic. It seems reasonable to assume that second-order Fermi acceleration caused by turbulence induced by any physical phenomena which supply sufficient energy to the \textit{in situ} relativistic electrons naturally explains the origin of these low-surface brightness bridges. 
A morphology comparison of the radio phoenix at the cluster periphery with the simulations presented by \citet{Enblin2002MNRAS.331.1011E} indicates its association with an outgoing shock.
In that case, the turbulence left behind by the outgoing shock naturally explains the bridge which is still luminous enough to be seen as connected to the phoenix. In Fig. \ref{fig:radio_xray}, we see the X-ray brightness contours, which can be treated as a proxy for the ICM density. Hence, the outgoing shock becomes stronger as it travels further from the cluster centre. 
This might explain the relatively smooth bridge closer to the cluster centre and the complex filamentary phoenix at the periphery in the wake of the same outgoing merging shock. 
However, the gradual increase in shock strength (with decreasing ICM density e.g., \citealt{Ha2018ApJ...857...26H}) fails to explain the brightness drop along the bridge and the abrupt jump in the radio brightness between the bridge and the phoenix, unless we consider the non-thermal electron population of the phoenix-forming cocoon to be much denser compared to the bridge region and possibly of different origin as well. 

Furthermore, as we have discussed earlier, the A85 ICM contains spiralling sloshing gas around the cluster centre.
We have shown a contour map of this spiralling gas in Fig. \ref{fig:uvsub_main} right panel. For this, an unsharp masked residual image was created after subtracting the best-fitted 2D-beta model from the X-ray surface brightness map. The two sub-clusters were excluded to prevent any biases during the model creation. A single beta model was found to be the best fit for the \textit{Chandra} image of the cluster (see \citealt{Rahaman2022MNRAS.515.2245R} for more details).
Thus, just like in the case of minihalos, the sloshing-induced turbulence can also be responsible for re-accelerating the bridge's non-thermal electron population. In Fig. \ref{fig:uvsub_main} (right) we see that the spiralling sloshing arm fully covers the bridge, which can also be a natural explanation for the origin of the bridge. However, the well-resolved Gaussian brightness structure across the bridge width distinguishes it from the surrounding minihalo. 
Furthermore, the azimuthal profile in Fig. \ref{fig:az_profile} shows that the bridge brightness is slightly higher than the outer part of the minihalo.
Thus, sloshing-driven turbulence alone fails to fully explain it and indicates the role of a possible other mechanism at play, which in this case could be the post-shock turbulence. Besides, although the uncertainties are large, the relatively flatter spectral index of the bridge ($-0.92$) compared to the literature values for bridges ($-1.3 \lesssim \alpha \lesssim -1.7$) hints that the particle energisation may be caused by more than one process,
i.e., the turbulence resulting from both the sloshing motion and the outgoing shock wave. 

The well-accepted explanation for the seed relativistic electrons in minihalos and phoenices is that they are of AGN origin. In the case of minihalos, the central AGN is the most likely candidate and the ICM sloshing is responsible for transport and distribution around the cluster centre \citep{Fujita2004ApJ...612L...9F,Mazzotta2008ApJ...675L...9M}. On the other hand, fossil relativistic electron bubbles from past AGN activity are considered to be the seed for radio phoenices \citep{Enblin2001A&A...366...26E,Enblin2002MNRAS.331.1011E}. Now, this begs the question, where do the bridge-producing electrons come from? 
The sharp edges in three directions of the phoenix indicate that the electrons responsible for the formation of this object are bound within the phoenix region. 
Furthermore, it is unlikely for these electrons to diffuse in the bridge direction only, where the ICM density increases with decreasing cluster-centric radius. Moreover, the rapid change in the brightness at the junction between the bridge and the phoenix further emphasises the possible electron density variation which is unlike the case of simple diffusion from the phoenix cocoon. In contrast, the sloshing activity might be responsible for non-thermal electron transport from the central AGN (similar to the minihalo), as the sloshing spiralling arm completely covers the bridge region (Fig. \ref{fig:uvsub_main} right panel). Furthermore, the decrease in the surface brightness along the bridge (Fig. \ref{fig:bridge_profile} right panel) may also indicate the transport of seed electrons from the centre to the outer regions by the sloshing motion. We would also like to point out that the spectral indices of the minihalo and the bridge are found to be similar. We speculate that this may as well point towards a similar origin of the seed relativistic electrons for these two diffuse objects.

\section{Conclusions} \label{sect:conc}
We have re-confirmed the discovery of the faint minihalo at the centre of the A85 cluster. 
The size of the minihalo at 700 MHz is found to be $\sim 280$ kpc. The spectral index between 700 MHz and 1.28 GHz is found to be $-0.97\pm0.29$, which is on the flatter side of the typical minihalo spectrum. The presence of the spiralling cool gas around the cluster centre supports the mechanism proposed by \citet{Fujita2004ApJ...612L...9F}.

We report the discovery of a radio bridge in A85, connecting the central minihalo and the peripheral radio phoenix.  
The average surface brightness of the $\sim 220$ kpc long bridge is $\sim 0.14\ \mu$Jy/arcsec$^2$ at 700 MHz, which is too faint to be detected in previous observations. The estimated integrated flux density is found to be $4.88\pm0.69$ and $2.8\pm0.59$ mJy at 700 and 1280 MHz, respectively. The spectral index between 700 MHz and 1.28 GHz is found to be $-0.92\pm 0.42$, which is much flatter than previously reported bridges. 
Both the shock wave responsible for the phoenix and the sloshing spiralling arm in the ICM may be responsible for the re-acceleration of the seed electrons. However, we do not yet know the origin of these seed electrons themselves. Further multi-frequency analysis is necessary for this endeavour.

\section*{Acknowledgements}
We would like to thank Rhodes University for providing the necessary computing facilities for data analysis. We thank the staff of GMRT, who made these observations possible. GMRT is run by the National Centre for Radio Astrophysics of the Tata Institute of Fundamental Research. 
The MeerKAT telescope is operated by the South African Radio Astronomy Observatory, which is a facility of the National Research Foundation, an agency of the Department of Science and Innovation. We would like to thank the referee Lawrence Rudnick for his valuable comments and suggestions which have greatly improved the quality of our work.
RR and OS's research is supported by the South African Research Chairs Initiative of the Department of Science and Technology and the National Research Foundation. MR acknowledges financial support from the Ministry of Science and Technology of Taiwan (MOST 109-2112-M-007-037-MY3). This research is supported by DST-SERB, through ECR/2017/001296 grant awarded to AD.

\section*{Data Availability}
The radio data used in this work are available in the GMRT Online Archive (\url{http://naps.ncra.tifr.res.in/goa/data/search}) and SARAO Web Archive (\url{https://archive.sarao.ac.za/}) with the proposal IDs given in the Sect. \ref{sect:obs}.



\bibliographystyle{mnras}
\bibliography{myBIB} 




\appendix

\section{Image local RMS and zero level} \label{sec:zero_level}
\begin{figure}
    \centering
    \includegraphics[width=\columnwidth]{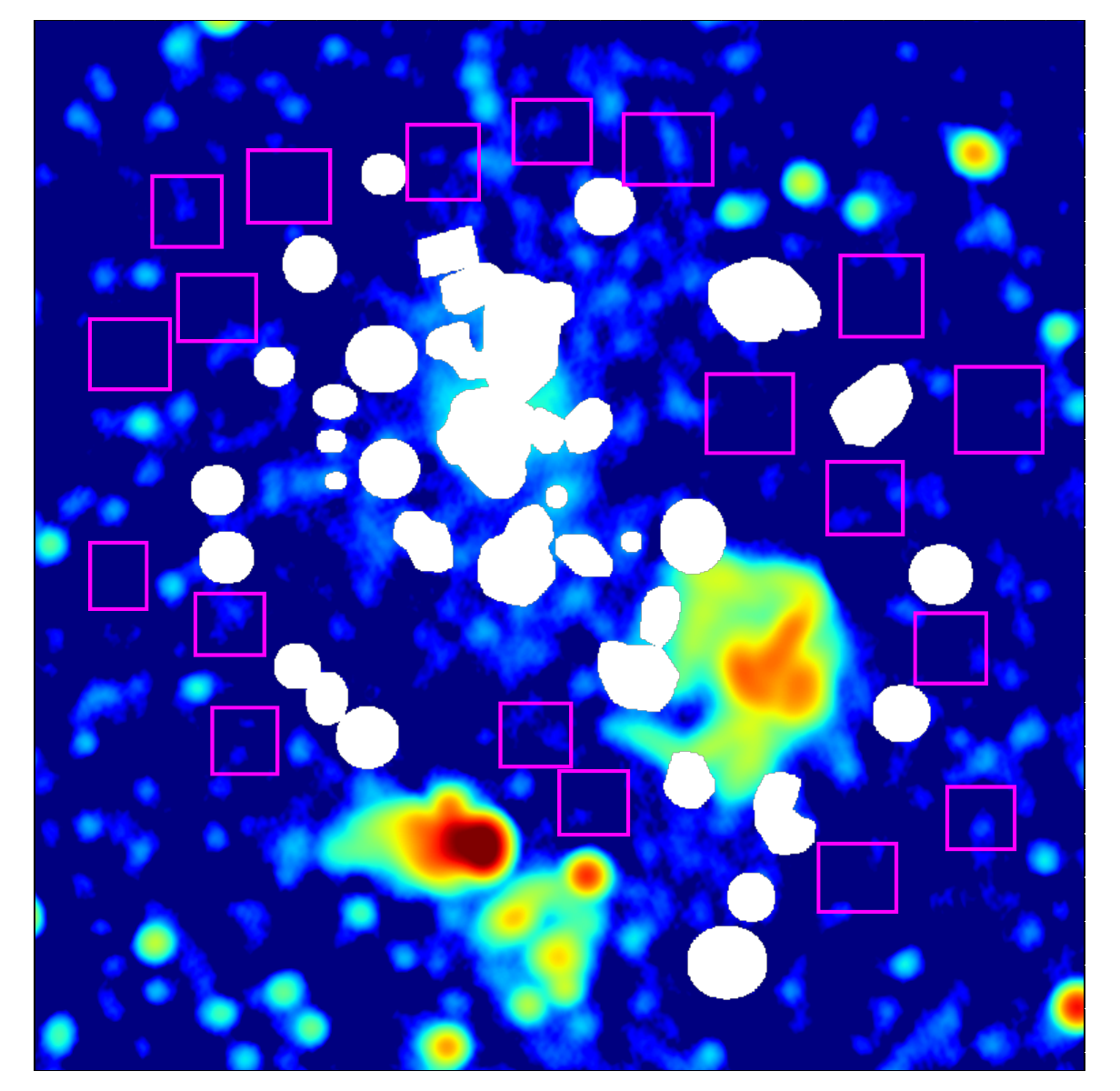}
    \caption{Regions around the target source to determine image `local RMS' and `zero level'.}
    \label{fig:zero_reg}
\end{figure}

To determine the `local RMS' noise and `zero level' around the target source we used the following method. First, we selected some source-free box regions (see Fig. \ref{fig:zero_reg}) around the target. Then, we calculated the mean, median and standard deviation values corresponding to each box region. After that, we took the average of the mean, median and std across the regions. This we take to be representative of the mean, median and std of the source-free regions around the target. Now, we assume `local RMS' to be the average `std'. We take the `zero level' to be the average median value across the regions. Furthermore, we take the scatter of the median values (zero level scatter or $zls$) across the regions as the uncertainty in the zero level estimate. After determining the `zero level', we subtracted this value from the brightness values of each pixel. This corrects for any `zero level' bias that may have been in the target region. Moreover, we also incorporated the scatter of the median values ($zls$) in our brightness and flux density error calculation. We present image statistics both before and after `zero level' correction for 700 and 1280 MHz images in Table \ref{tab:zero700} \& \ref{tab:zero1280}, respectively.

\begin{table}
\centering
\scalebox{0.75}{
\begin{tabular}{|c|c|c|c|c|c|c}
\hline
Label  & Avg\_Mean & Avg\_Median & Avg\_Std & Local RMS & Zero level & zls\\ \hline
Before & -4.07e-06 & -2.56e-06   & 2.63e-05 & 2.63e-05  & -2.56e-06 & 1.51e-05 \\ \hline
After  & -1.52e-06 & -3.83e-13    & 2.63e-05 & 2.63e-05  & -3.83e-13 & 1.51e-05  \\ \hline
\end{tabular}}
\caption{700 MHz image statistics before and after zero level correction.}
\label{tab:zero700}
\end{table}

\begin{table}
\scalebox{0.75}{
\begin{tabular}{|c|c|c|c|c|c|c}
\hline
Label  & Avg\_Mean & Avg\_Median & Avg\_Std & Local RMS & Zero level & zls\\ \hline
Before & 1.26e-05  & 1.06e-05    & 2.21e-05 & 2.21e-05  & 1.06e-05 & 1.37e-05  \\ \hline
After  & 1.98e-06  & -7.66e-13   & 2.21e-05 & 2.21e-05  & -7.66e-13 & 1.37e-05 \\ \hline
\end{tabular}}
\caption{1280 MHz image statistics before and after zero level correction.}
\label{tab:zero1280}
\end{table}

\section{UV-sampling at short baselines}
\begin{figure*}
    \centering
    \begin{tabular}{cc}
     \includegraphics[width=0.99\columnwidth]{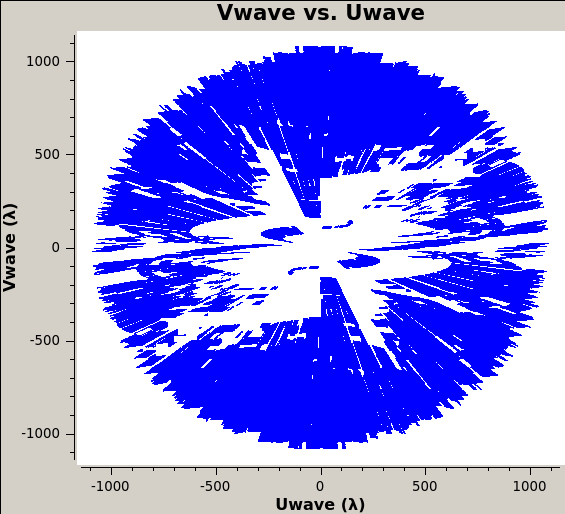}    & \includegraphics[width=0.99\columnwidth]{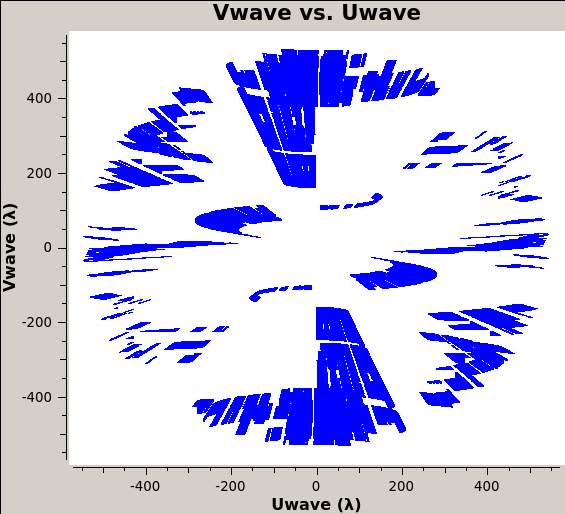} \\
     \includegraphics[width=0.99\columnwidth]{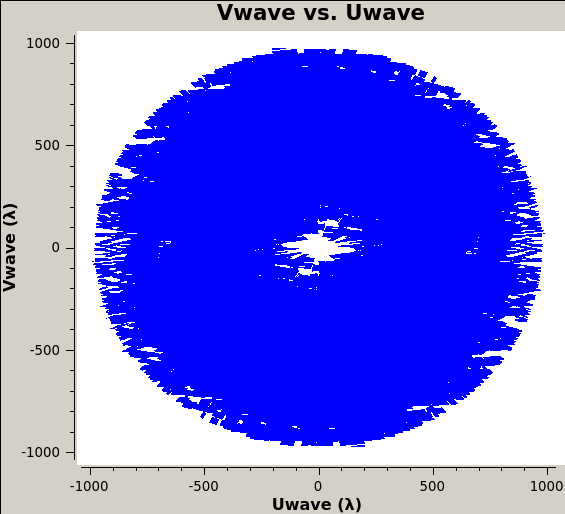}   & \includegraphics[width=0.99\columnwidth]{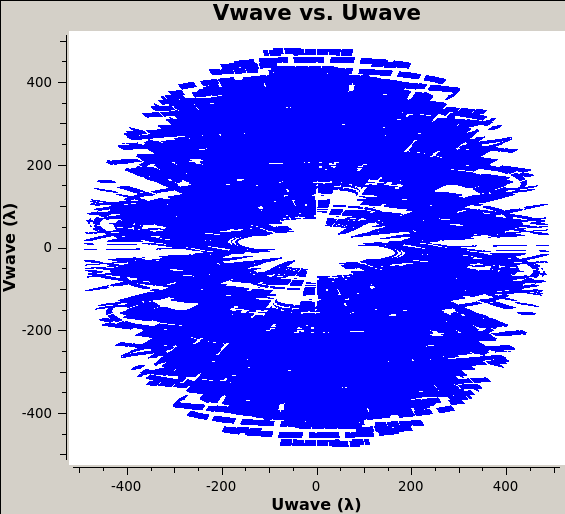}
    \end{tabular}
    \caption{\textit{uv}-plot of the short baselines at 700 MHz (top row) and 1.28 GHz (bottom row) observations.}
    \label{fig:uvplot}
\end{figure*}


\bsp	
\label{lastpage}
\end{document}